\documentclass[aps,prb,twocolumn,footinbib,superscriptaddress,floatfix,showkeys]{revtex4}

\usepackage{color}
\usepackage{epsfig}
\usepackage{latexsym}
\usepackage{amssymb}
\usepackage{amsmath}
\usepackage{algorithm}
\usepackage{algorithmic}
\usepackage{latexsym}
\usepackage{fancyhdr}
\usepackage{verbatim}
\usepackage{tabularx}
\usepackage{epsfig}
\usepackage{amsmath}
\usepackage{amssymb}
\usepackage{graphicx}
\usepackage{wasysym}
\usepackage{subfigure}
\usepackage{setspace}
\usepackage{xr}

\usepackage{tgheros}

\usepackage[T1]{fontenc}

\definecolor{bv}{RGB}{52,43,125}

\begin{document}

\title{Limitations in Predicting the Space Radiation Health Risk for Exploration Astronauts}

\author{\color{bv}{Jeffery C. Chancellor}}
\affiliation{Department of Physics and Astronomy, Texas A\&M University, College Station, TX 77843-4242, USA}

\author{\color{bv}{Rebecca S. Blue}}
\affiliation{Aerospace Medicine and Vestibular Research Laboratory, The Mayo Clinic Arizona, Scottsdale, AZ 85054, USA}

\author{\color{bv}{Keith A. Cengel}}
\affiliation{Department of Radiation Oncology, Perelman School of Medicine, University of Pennsylvania, Philadelphia, PA 19104, USA}

\author{\color{bv}{Serena M. Au\~n\'on-Chancellor}}
\affiliation{University of Texas Medical Branch, Galveston, TX 77555, USA}
\affiliation{National Aeronautics and Space Administration (NASA), Johnson Space Center, Houston, TX 77058, USA}

\author{\color{bv}{Kathleen H. Rubins}}
\affiliation{National Aeronautics and Space Administration (NASA), Johnson Space Center, Houston, TX 77058, USA}

\author{\color{bv}{Helmut G.~Katzgraber}}
\affiliation {Department of Physics and Astronomy, Texas A\&M University, College Station, TX 77843-4242, USA}
\affiliation{1QB Information Technologies (1QBit), Vancouver, British Columbia, V6B 4W4, Canada}
\affiliation {Santa Fe Institute, 1399 Hyde Park Road, Santa Fe, NM 87501, USA}

\author{\color{bv}{Ann R. Kennedy}}
\affiliation{Department of Radiation Oncology, Perelman School of Medicine, University of Pennsylvania, Philadelphia, PA 19104, USA}


\maketitle

{\color{bv}{\bf Despite years of research, understanding of the space radiation environment and the risk it poses to long-duration astronauts remains limited. There is a disparity between research results and observed empirical effects seen in human astronaut crews, likely due  to the numerous factors that limit terrestrial simulation of the complex space environment and extrapolation of human clinical consequences from varied animal models. Given the intended future of human spaceflight, with efforts now to rapidly expand capabilities for human missions to the moon and Mars, there is a pressing need to improve upon the understanding of the space radiation risk, predict likely clinical outcomes of interplanetary radiation exposure, and develop appropriate and effective mitigation strategies for future missions. To achieve this goal, the space radiation and aerospace community must recognize the historical limitations of radiation research and how such limitations could be addressed in future research endeavors. We have sought to highlight the numerous factors that limit understanding of the risk of space radiation for human crews and to identify ways in which these limitations could be addressed for improved understanding and appropriate risk posture regarding future human spaceflight.}}

\section*{Introduction}
While space radiation research has expanded rapidly in recent years, large uncertainties remain in predicting and extrapolating biological responses to radiation exposure in humans. As future missions explore outside of \emph{low-Earth orbit} (LEO) and away from the protection of the Earth's magnetic shielding, the nature of the radiation exposures that astronauts encounter will include higher radiation exposures than any experienced in historical human spaceflight. In 1988, the \emph{National Council on Radiation Protection and Measurements} (NCRP) released Report No.~98:~Guidance On Radiation Received in Space Activities.\cite{ncrp_guidance_1989} In this report, authors recommended that NASA astronauts be limited to career lifetime radiation exposures that would induce no more than a 3\% \emph{Risk of Exposure-Induced Death} (REID). This was re-emphasized in the 2015 NCRP Commentary No.~23:~Radiation Protection for Space Activities: Supplement to Previous Recommendations, which concluded that NASA should continue to observe the 3\% REID career limit for future missions outside of LEO.\cite{national_council_on_radiation_protection_and_measurements_ncrp_radiation_2015} This limit has been accepted in NASA's Spaceflight Human-System Standard document, NASA STD-3001 Volume 1 (Revision A).\cite{NASASTD3001}

Despite the adoption of these guidelines and the past 30 years of research, there has been little progress on fully defining or mitigating the space radiation risk to human crew. In fact, the NCRP's recent conclusions specified that their 3\%\ limit may not be conservative enough given the incomplete biological data used in existing projection models, and that such models may overestimate the number of allowable "safe days" in space for missions outside of LEO.\cite{national_council_on_radiation_protection_and_measurements_ncrp_radiation_2015} As a result, NASA has yet to establish mission exposure limits for crews of exploration-class missions outside of LEO.

A recent report by Schwadron \emph{et~al.}~has identified further concerns regarding the interplanetary radiation environment.\cite{schwadron_does_2014} The unusually low activity between solar cycles 23 and 24 (1996-present) has resulted in the longest period of minimum solar activity observed in over 80 years of solar measurements. The lack of solar activity has led to a substantial decrease in solar wind density and magnetic field strengths that typically attenuate the fluence (the flux of particles crossing a given plane) of \emph{Galactic Cosmic Ray} (GCR) ions during periods of solar minimum. As a result, Schwadron \emph{et al.}~project that GCR fluences will be substantially higher during the next solar cycles (24-25) leading to increased background radiation exposure and, subsequently, as much as a 20\% decrease in the allowable safe days in space (outside of LEO) to stay below the 3\% REID limits.\cite{schwadron_does_2014}

\begin{figure}[ht]
\centering
\includegraphics[width=\linewidth,keepaspectratio]{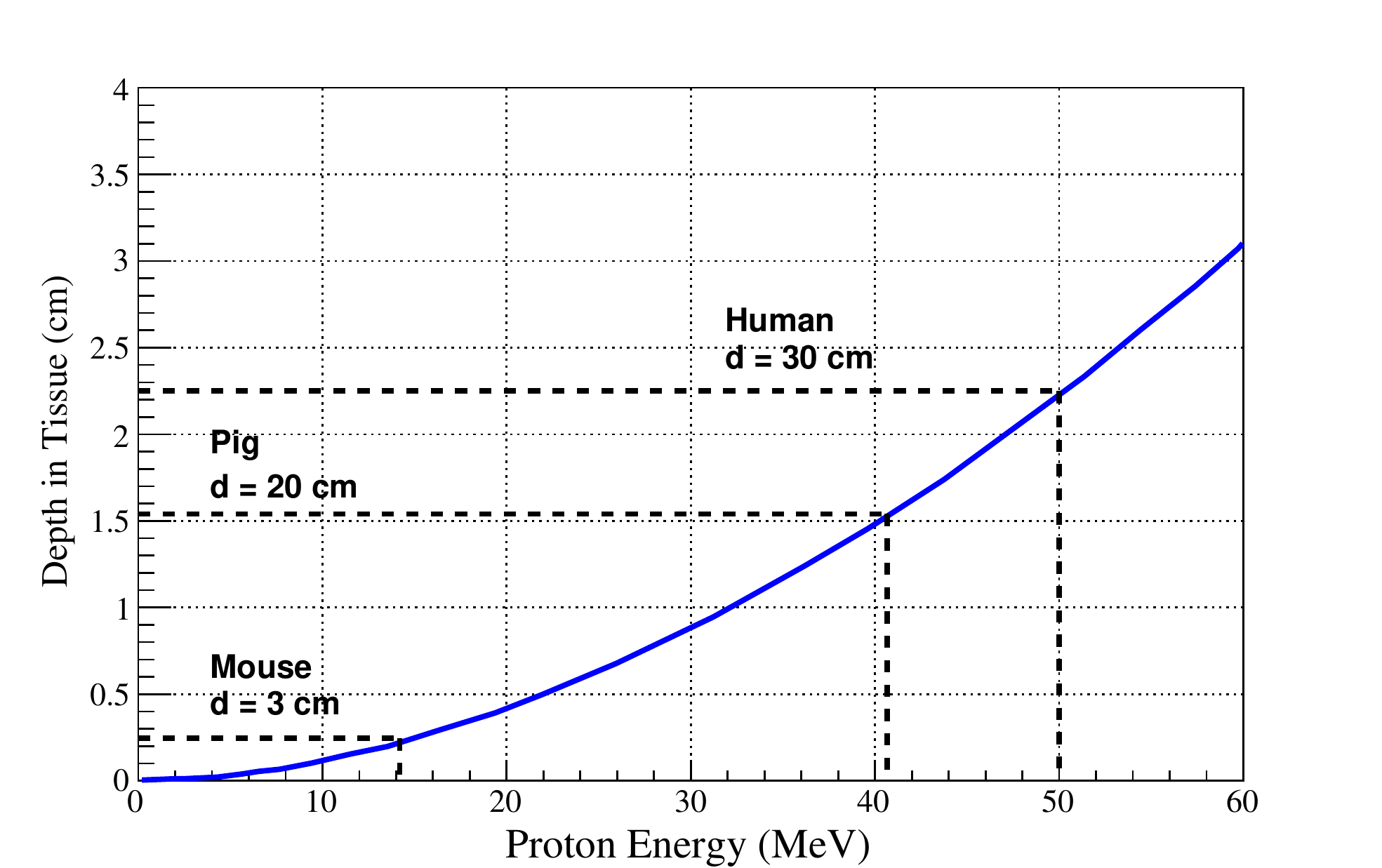}
\vspace*{0.5em}
\caption{Depth dose, energy, and linear energy transfer characteristics of protons. The range of proton energies relative to the body diameter (dotted lines) and bone marrow depth (ordinate) for mice, pigs, and humans for energies up to 60MeV.}
\label{fig:range}
\end{figure}

The study of human health risks of spaceflight (e.g. bone health, behavior, nutrition, etc.) typically involves analogs that closely represent the space environment. In most cases, theory, models, and study outcomes can be validated with available spaceflight data or, at a minimum, observation of humans subjected to analog terrestrial stresses. In contrast, space radiation research is limited to the use of analogs or models that for many reasons do not accurately represent the operational space radiation environment or the complexity of human physiology. For example, studies on the effects of space radiation generally use mono-energetic beams and acute, single-ion exposures (including protons, lithium, carbon, oxygen, silicon, iron, etc.) instead of the complex energy spectra and diverse ionic composition of the space radiation environment. In addition, a projected, cumulative mission dose is often delivered in one-time, or rapid and sequential, doses delivered to experimental animals. In most cases, these dose-rates are several orders of magnitude higher than actual space environment exposures. Even the use of animal models introduces error, as studies make use of a variety of animal species with differing responses and sensitivity to radiation that may not represent human responses to similar exposures. Further, studies do not challenge multiple organ systems to respond concurrently to the numerous stressors seen in an operational spaceflight scenario. Historical epidemiological studies of humans, which are generally used for correlation of animal and experimental models, include populations such as atomic bomb or nuclear accident survivors exposed to whole-body irradiation at high doses and high dose-rates, limited to scenarios not found in spaceflight. These disparities and numerous other environmental considerations contribute to the large uncertainties in the outcomes of space radiobiology studies and the applicability of such studies for extrapolation and prediction of clinical health outcomes in future spaceflight crews.

Here we seek to highlight these factors that contribute to the challenge of radiation risk prediction and mitigation for future exploration spaceflight. Our intent is to provide an understanding of the current state of radiation-specific literature, efforts towards better defining the space radiation environment, and the difficulties in realization of this effort that limit current knowledge. Further, we hope to identify opportunities for future research that could best elucidate a path towards successful definition and mitigation of the space radiation risk to humans outside of LEO.

\section*{The Space Radiation Environment}

Biological stressors related to space radiation are due to the effects of energy transfer from a charged particle to the human body. The combination of a particle's charge, mass, and energy determines how quickly it loses energy when interacting with matter.\cite{attix_introduction_1986,hall_radiobiology_2012,ziegler_handbook_2013} For example, given equal initial kinetic energies, an electron will penetrate further into aluminum than a heavy charged particle, and an x-ray will, on average, penetrate even further. In biological tissue, the absorbed dose that a particular target organ receives from heavy-charged particle radiation depends not only on the energy spectrum of the particles but also on the depth and density of the tissue mass that lie between the skin surface and the target organ (for example, see Figure \ref{fig:range}, which demonstrates the tissue depth ionized hydrogen (proton) penetrates as a function of energy).

The radiation dose to an astronaut, measured in units of \emph{Gray} (\emph{Gy}, defined as \emph{Joules} per kilogram (\emph{$J/kg$})), is deposited with a distribution in tissues that results from the specific energy fluence of the particles. The heavier the charged particle, the greater the amount of energy deposited per unit path length for that particle. This is called linear energy transfer (LET).

\begin{figure*}[!ht]
\centering
\includegraphics[width=\linewidth,height=8cm,keepaspectratio]{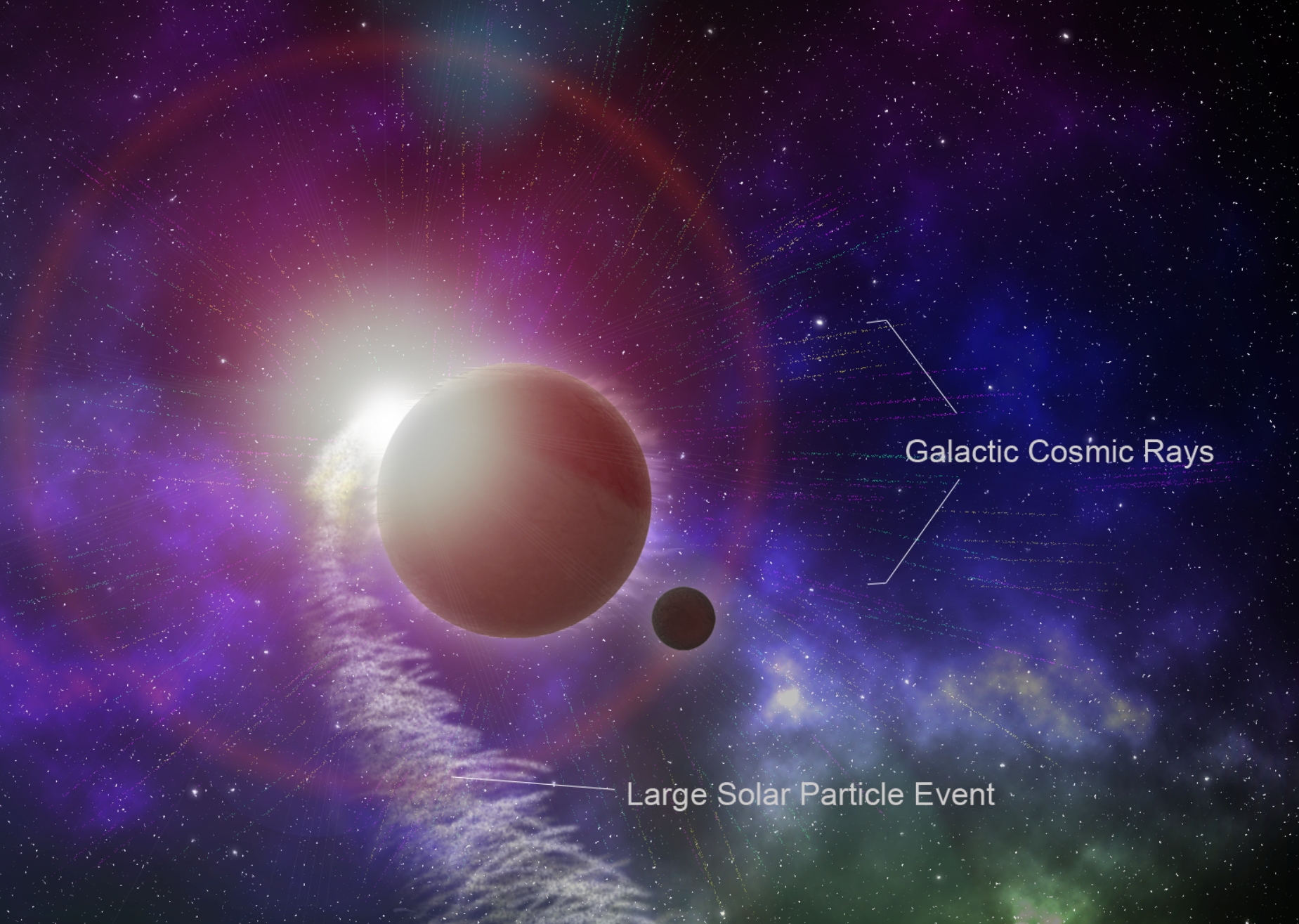}
\vspace*{0.5em}
\caption{Interplanetary Radiation Environment. Galactic Cosmic Rays (GCR) and unpredictable Solar Particle Events (SPEs) pose a significant threat to astronauts during exploration missions to the moon or Mars. Each radiation source has a unique impact on tissue health, shielding design, and mitigation strategies during spaceflight operations. We are uncertain of how simultaneous and prolonged exposure to these radiations will affect short- or long-term human health. Illustration by R.~Blue.}
\label{fig:environment}
\end{figure*}

The space weather environment is most commonly categorized into three sources of ionizing radiation, each of which is associated with different energy and prevalence and, thus, different radiation-related risk (Figure 2). First, the GCR spectrum consists of primarily ionized hydrogen as well as less frequent heavier-charged particles, with relatively high LET, that contribute to the chronic, background radiation exposure for long-duration astronauts. \emph{Solar Particle Events} (SPEs) consist mostly of short-duration exposures of high-energy protons that emanate from the Sun within regions of solar magnetic instability.\cite{national_council_on_radiation_protection_information_2006} Finally, solar wind consists of mostly low energy protons and electrons. The background dose-rate for solar wind varies with the solar cycle, but is easily shielded by modern spacecraft designs and is considered to be of negligible risk. In addition to space environment radiation, some small amounts of radioisotopes are used in manned space missions for instrument calibration and research; however, these sources are highly controlled by flight rules and mission planners. The vast majority of crew radiation exposures are delivered by the complex radiation environment in which they must travel and live.

\subsubsection*{Galactic Cosmic Rays}
\begin{figure}[ht]
\centering
\includegraphics[width=\linewidth,keepaspectratio]{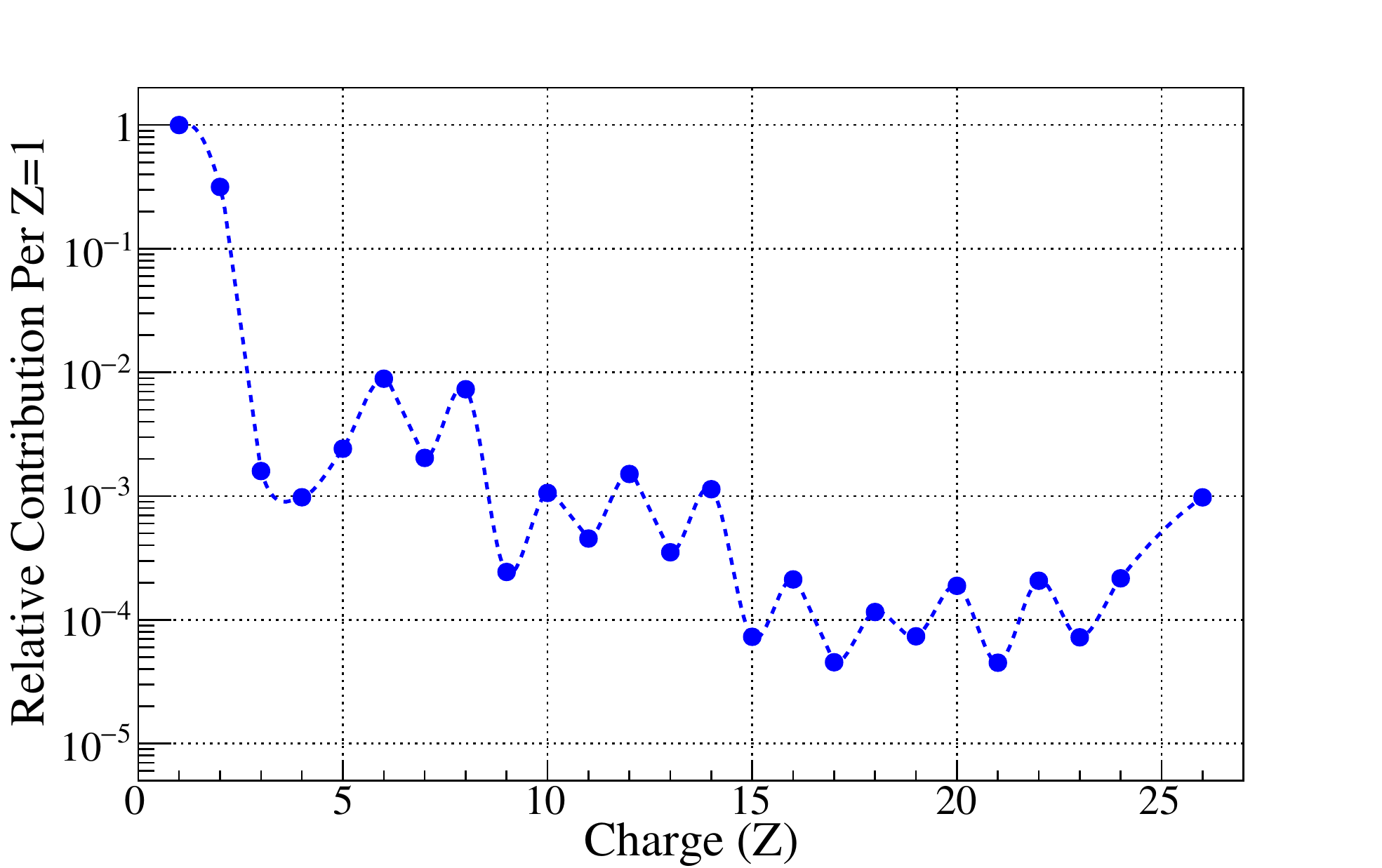}
\vspace*{0.5em}
\caption{Relative abundance of atomic species, normalized to Z=1 (hydrogen) and up to Z=26 (iron), in the Galactic Cosmic Ray (GCR) spectrum. The GCR spectrum includes every atom in the periodic table, with ions up to nickel (Z=28) contributing to any significance. Note the energy of each ion species varies widely, more prominently in the range of 400-600MeV. This broad disparity in ions and energies makes it extremely difficult to accurately simulate the GCR environment during ground-based radiobiology experiments. While larger ions may provide lower relative contribution to the spectrum makeup they may have a more significant biological impact than smaller, abundant ions. Figure adapted from Simpson \emph{et al.}~1983.\cite{simpson_elemental_1983}}
\label{fig:GCR}
\end{figure}

GCR ions, originating from outside our solar system, are relativistic nuclei that possess sufficient energies to penetrate any shielding technology used on current mission vehicles.\cite{cucinotta_evaluating_2006} The GCR spectrum is a complex combination of fast-moving ions derived from most atomic species found in the periodic table.\cite{simpson_elemental_1983} The GCR spectrum, from hydrogen (Z, or atomic number, of 1) through iron (Z=26), is shown in Figure \ref{fig:GCR}. This spectrum consists of approximately $87\%$ hydrogen ions (protons), $12\%$ helium ions ($\alpha$ particles), and 1-2\% heavier nuclei with charges ranging from Z=3 (lithium) to Z=28 (nickel).\cite{simpson_elemental_1983,badhwar_radiation_1998} Ions heavier than nickel are also present, but they are rare in occurrence. GCR ions with charge $Z\geq3$ are frequently referred to as \emph{HZE} particles (\textit{H}igh nuclear charge \emph{Z} and energy \emph{E}).

During transit outside of LEO, every cell nucleus within an astronaut's body would be traversed by a hydrogen ion or delta ray (a recoil electron caused by fragmentation after ion interactions) every few days, and by a heavier GCR ion (\emph{e.g.} O, Si, Fe) every few months.\cite{cucinotta_space_2001,cucinotta_cancer_2006} Despite their infrequency, the heavy ions contribute a significant amount to the GCR dose that astronauts would incur outside of LEO. The energies of the heavier GCR ions are so penetrating that shielding can only partially reduce the intravehicular doses.\cite{cucinotta_cancer_2006} Thicker shielding could provide protection, but is limited by mass and volume restrictions of exploration vehicles and dependent upon the capabilities of spacecraft launch systems.

The high-LET radiation found in the GCR spectrum can produce excessive free radicals that instigate oxidative damage to cell structures. Chronic exposure to such oxidative stress contributes to the radiation-induced changes associated with premature aging, cardiovascular disease, and the formation of cataracts. The large ionization power of GCR ions makes them a potentially significant contributor to tissue damage and carcinogenesis, \emph{central nervous system} (CNS) degeneration, and deleterious health outcomes.\cite{chancellor_space_2014,walker_heavy_2013} In addition, as GCR ions pass through a space vehicle, interaction with the spacecraft hull attenuates the energy of heavy-charged particles and frequently causes their fragmentation into numerous particles of reduced atomic weight, a process referred to as \textit{spallation}.\cite{rossi1952high,hodgson1997introductory} Spallation occurring as GCR particles collide with shielding materials can result in `cascade showers' that produce progeny ions with much higher potential for biological destruction than the original particle.\cite{cucinotta_evaluating_2006,cucinotta_cancer_2006,guetersloh_polyethylene_2006,townsend_estimates_1994} This process changes the makeup of the intravehicular radiation spectrum, adding to the complexity of the radiation environment unique to spaceflight.

\subsubsection*{Solar Particle Events}

During SPEs, magnetic disturbances on the surface of the sun result in the release of intense bursts of ionizing radiation that are difficult to forecast in advance.\cite{hellweg_getting_2007,wilson_shielding_1999,smart_comment_2003} SPE radiation is primarily composed of protons with kinetic energies ranging from 10MeV up to several GeV (determined by the relativistic speed of particles) and is predicted to produce a heterogeneous dose distribution within an exposed astronaut's body, with a relatively high superficial (skin) dose and a significantly lower dose to internal organs.

As extravehicular space suits provide relatively low shielding protection, SPE exposures occurring during extravehicular activities would pose significant risk to astronauts.\cite{hu_modeling_2009} However, astronauts would still receive potentially significant elevations in radiation dose even within a shielded spacecraft and remain vulnerable, especially on long-duration missions, to both acute effects of sudden SPE radiation boluses and to the overall additive effects of GCR and repetitive SPEs over the course of a mission.

While many SPEs show modest energy distributions, there are occasional and unpredictable high fluence events; for example, a particularly large SPE in October 1989 is predicted to have delivered dose-rates as high as 1,454mGy/hour to an exposed astronaut in a vehicle traveling in interplanetary space (for context, consider that the daily dose for long-duration astronauts aboard the ISS is approximately 0.282mGy per day).\cite{hu_modeling_2009,wilson_2000,GSFC}  Similarly, some SPE can deliver particularly high-energy doses: for example, 10-15\% of the total fluence of an October 1989 SPE was made up of protons with energies in excess of 100MeV.\cite{ncrp_guidance_1989,hu_modeling_2009} If an astronaut were exposed to such an event during long-duration spaceflight, there are potential risks for both acute radiation-induced illnesses and for significant increase in the overall mission dose accumulation. It should be noted that these predictions made use of classic shielding values (5g/cm$^{2}$) similar to those of the Apollo command module (average shielding of 6.15g/cm$^{2}$).\cite{clowdsley2005radiation}

Energetic SPE events produce protons with energies $\geq$ 100MeV that would penetrate classic spacecraft shielding, potentially reaching blood-forming organ depths with deleterious clinical sequelae. These highly energetic SPE exposures delivered to crews undertaking interplanetary flight could result in potentially serious symptoms ranging from prodromal responses (nausea, vomiting, fatigue, weakness) to fatality. In addition, large SPE doses can produce degenerative effects associated with cancer, ocular cataracts, respiratory and digestive diseases, and damage to the microvasculature; while these effects are mostly latent and do not necessarily pose an immediate risk to crew health, their overall impact upon long-duration crews is an important consideration.\cite{wu_risk_2008}

\subsubsection*{Interplanetary Radiation Environment}

The fluence of GCR particles in interplanetary space fluctuates inversely with the solar cycle, with dose-rates of 50-100mGy/year at solar maximum to 150-300mGy/year at solar minimum.\cite{mewaldt_cosmic_2005} The fluence and occurrence of SPEs is unpredictable, but dose-rates as high as 1,400--2,837mGy/hour are possible.\cite{ncrp_guidance_1989,national_council_on_radiation_protection_information_2006,hu_modeling_2009}

\begin{figure*}[!ht]
\vspace*{-2.5em}
  \begin{center}
    \subfigure[]{\label{fig:depthdose}\includegraphics[scale=0.7]{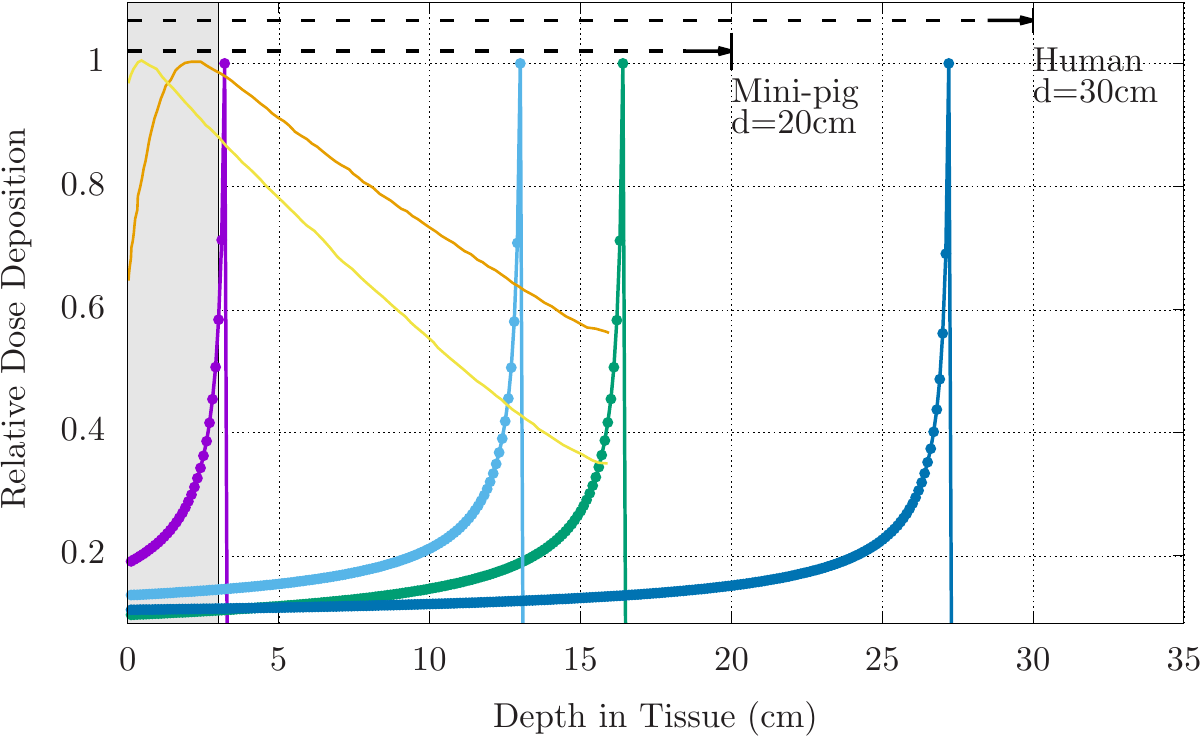}}
    \subfigure[]{\label{fig:89SPE}\includegraphics[scale=0.7]{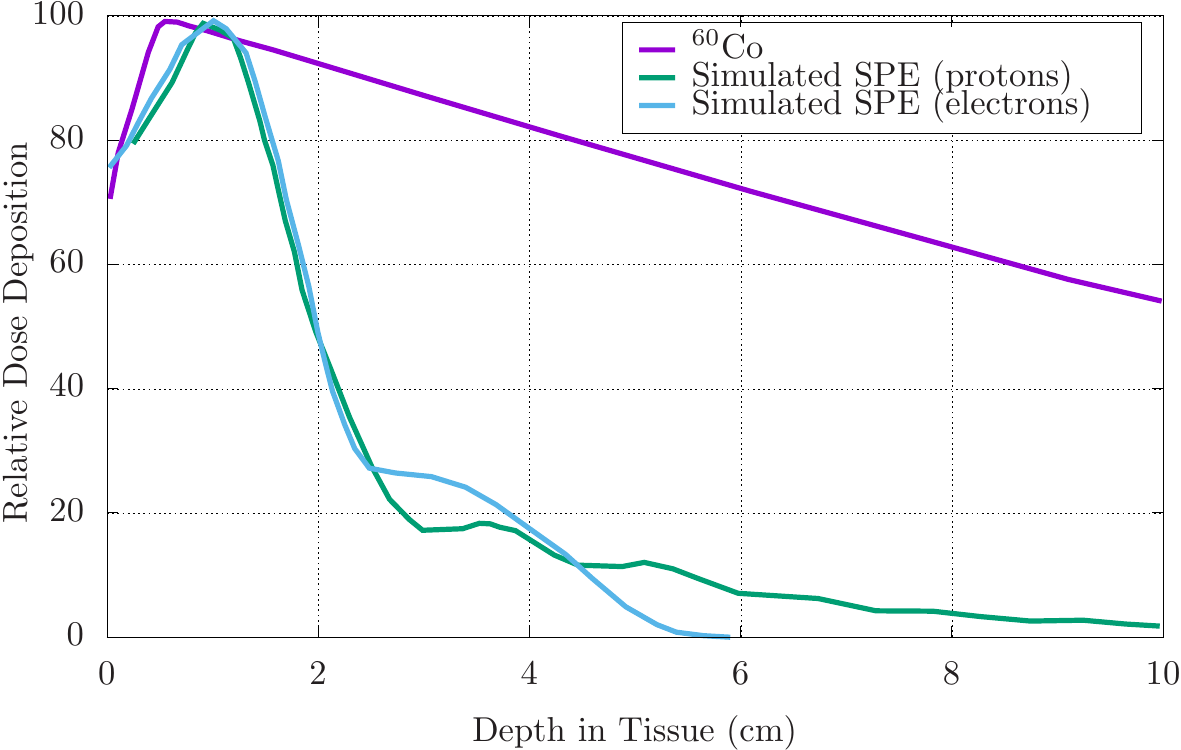}}
  \end{center}

\caption{(a): The Bragg peak and depth dose characteristics of space radiation. The Bragg peak and relative dose
deposition for ions at energies commonly used in space radiation studies compared to the x-ray and gamma sources used as
surrogate radiations for Relative Biological Effectiveness (RBE) quantification. The Bragg peak refers to the point where
a charged particle promptly loses kinetic energy before coming to rest in a medium. This effect is very pronounced for
fast moving, charged particles. Shown are 60MeV Protons (hydrogen, purple), 600MeV $^{56}$Fe (iron, light blue),
290MeV$^{12}$C (carbon, green), 1GeV $^{56}$Fe (iron, dark blue), x-ray (orange dotted line), and $^{60}$Co (cobalt,
yellow dotted line). The shaded gray area, representing the average diameter of a mouse, demonstrates that the Bragg peak, and thus the majority of dose deposition, is \emph{outside} the mouse body for SPE protons (energies $\geq$50MeV) and GCR ions. (b): The proton and electron range, energy and dose distributions for the October 1989 solar particle event compared to a dose-equivalent $^{60}$Co exposure. Charged particles (electrons, protons, heavy-charged particles) typically deposit more energy towards the end of their range. In contrast, the current standard, $^{60}$Co radiation, loses the most energy at the tissue surface. These energy characteristics demonstrate the poor fidelity of $^{60}$Co as a surrogate for studying the complex SPE and GCR spectrums. Adapted from Cengel \emph{et~al.}~2010 \cite{cengel_using_2010}.}

\label{fig:bragg_LET}
\end{figure*}

As discussed above, even if shielding in spacecraft effectively reduces radiation dose to the crew from SPEs, spallation occurring as GCR particles collide with shielding materials may lead to biological damage.\cite{cucinotta_evaluating_2006,cucinotta_cancer_2006,guetersloh_polyethylene_2006,townsend_estimates_1994} Aluminum shielding greater than 20-30g/cm$^{2}$ could only reduce the GCR effective dose by no more than 25\%.\cite{cucinotta_managing_2005} An equivalent mass of polyethylene would only provide about a 35\% reduction in GCR dose.\cite{edwards_rbe_2001,setlow_radiation_1996} While this degree of shielding has been achieved aboard the International Space Station (ISS), similar shielding is impractical within exploration mission design parameters due to the limited lift-mass capabilities of planned space launch systems. The Apollo crew module is the only vehicle to date that has transported humans outside of LEO; this vehicle could only effectively shield SPE protons with energies $\leq$75MeV. \cite{clowdsley2005radiation} To date, no studies have successfully emulated the complexity of energetic elements of the intravehicular radiation spectrum that astronauts are actually exposed to during space travel, or successfully incorporated vehicular design and shielding parameters in analog testing environments, limiting the understanding of the true effects of such an environment on the human body.

\section*{Challenges in Estimating Radiobiological Effect}

\subsection*{Modeling the Transfer of Energy}

As a charged particle traverses a material (such as spacecraft shielding, biological tissue, etc.), it continuously loses energy in particle interactions until the particle escapes the medium or has slowed enough to have strong interactions with orbiting electrons. This results in a rapid loss of particle energy over a very small distance with a corresponding rapid and sharp rise in LET. The `Bragg peak' (Figure \ref{fig:bragg_LET}(a)) describes the rapid transfer of kinetic energy from a charged particle before the particle comes to rest in a medium. This peak is particularly pronounced for fast-moving, charged particles, indicating more substantial energy transfer and, as a result, the potential for greater deleterious biological effect from such particles. However, if a particle instead passes directly through tissue without sufficient energy loss to provide effective stopping power, the sudden energy loss associated with a Bragg peak does not occur and damage is minimal. Space radiation studies to date generally presume a homogeneous distribution of energy loss inclusive of the Bragg peak for each type of radiation, likely overestimating the relative damage of some exposures.\cite{cengel_using_2010} Improved modeling of dose deposition and resultant biological sequelae specific to the space environment would advance risk estimation capabilities.

The biological effects of space radiation depend on multiple particle- and energy-specific factors, such as the LET specific to each ion, as well as the dose-rate of exposure. The \emph{Relative Biological Effectiveness} (RBE) of a particular radiation type is the numerical expression of the relative amount of damage that a fixed dose of that type of radiation will have on biological tissues. Higher RBEs are associated with more damaging radiation for a given dose. RBE is determined using the effectiveness of cobalt ($^{60}$Co) gamma rays as a standard. An RBE$=$1 means that the "test" radiation type (for example, heavy ion exposure) is as effective as $^{60}$Co radiation at producing a biological effect, and an RBE\textgreater1 means that the test radiation is more effective than $^{60}$Co radiation at producing a biological effect. However, in some cases this comparative value does not fully represent the energy transfer curve of a specific radiobiological insult (Figure \ref{fig:89SPE}).

The effect of quantifying factors such as LET, particle identity, dose-rate, and total dose on RBE remains incompletely understood. The RBE can vary for the same particle type, depending on energy, dose-rate, target organ, and other factors. Different particle types are assigned a radiation \textit{weighting factor} (formerly \textit{quality factor}), $W_{\rm R}$, that represents an average of calculated RBEs for a given particle. To identify the relative biological risk of a specific type and dose of radiation exposure, the physical dose (in Gy) is multiplied by $W_{\rm R}$ to obtain the biologically effective dose in units of \emph{Sieverts} ($Sv$). This method of estimating dose and relative effect introduces limitations in predicting the true biological risk of exposures, particularly exposures to complex and poorly understood radiation environments.

\subsection*{Limitations of Terrestrial Analogs}

\subsubsection*{Mechanisms of Biological Impact}

There are numerous limitations of current terrestrial analogs used for studying and predicting space radiation effects on biological tissues. The
mechanisms that cause biological damage from space radiation are uniquely different from those associated with terrestrial radiation sources that are frequently
used as surrogates in space radiobiology studies. Charged particle radiation, including GCR and SPE, causes primarily \emph{direct ionization} events, where biological effects are the direct result of interactions between the charged ion and impacted tissue. As charged particles lose energy successively through material interactions, each energy loss event can result in damage to the biological tissue. In contrast, terrestrial analogs often use radiation that causes \emph{indirect ionizing} events. In indirect ionization, non-charged particles, such as photons, interact with other molecules and cause the release of charged particles, such as free radicals or electrons, that ultimately cause biological damage. Thus, it is difficult to extract a
meaningful estimation of the direct ionizing space radiation impact through the use of terrestrial analogs and indirect ionizing radiation.

\begin{figure}[ht]
\centering
\includegraphics[width=\linewidth,keepaspectratio]{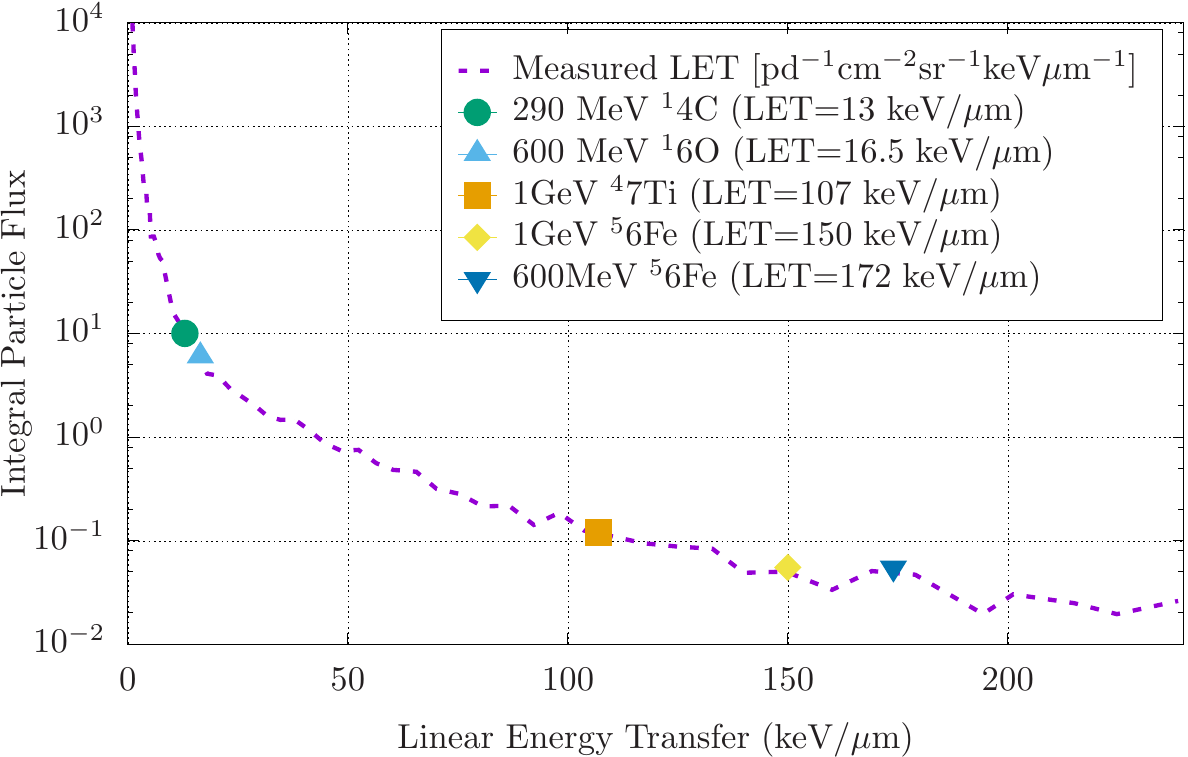}
\vspace*{0.5em}
\caption{Simulated gamma, proton, electron ranges in tissue. Displayed are the integrated LET/day values measured by Badhwar \emph{et al.}~1998 (purple dotted line) \cite{badhwar_radiation_1998}, as well as the LET of five single-ion exposures (290MeV $^{14}$C (carbon), 600MeV $^{16}$O (oxygen), 1GeV $^{47}$Ti (titanium), 1GeV $^{56}$Fe (iron), and 600MeV $^{56}$Fe (iron)). As studies generally focus on a single, mono-energetic radiation exposure, this figure highlights the lack in breadth of energies or radiation field complexity used in current radiobiological studies.}
\label{fig:LET}
\end{figure}

\subsubsection*{Cumulative Dose Delivery and Tissue Distribution}

Models of the space environment outside of LEO have predicted that astronaut crews may receive a total body dose of approximately 1-2mSv/day in interplanetary space and approximately 0.5-1mSv/day on the Martian surface.\cite{cucinotta_cancer_2006,saganti_radiation_2004} These doses would increase with any SPE encountered over the course of the mission.

Many recent studies have led to ominous conclusions regarding the non-acute effects of GCR radiation on CNS and cardiovascular health that are difficult to interpret as real effects likely to occur in humans, but suggest that the protracted, low dose and dose-rate radiation exposure expected on the longer, exploration missions might lead to mission-relevant threats to astronaut health.\cite{cherry_galactic_2012,parihar_what_2015} These experiments were performed using rodent models exposed to single ion, mono-energetic heavy-ion beams, in some cases at total doses that are many times higher than the radiation human crews would experience during interplanetary space travel.\cite{Durante2002,Durante2005,Wang2014} Even in studies where lower total doses are used, study methods delivered the cumulative mission doses for an entire mission over a very short period of time, typically over a few minutes.\cite{norbury_galactic_2016,Loucas2015} These parameters do not allow for critical physiologic components of the radiobiological response that would be expected under chronic, low-dose and low-dose-rate radiation conditions, such as cell regrowth and up-regulation of repair mechanisms.\cite{hall_radiobiology_2012} Additionally, there is substantial evidence that GCR exposure at the dose-rates expected in interplanetary space may not induce acute or subacute biological responses, while acute exposure to total/cumulative dosage easily could.\cite{wu_risk_2008}

Recently, NASA has developed an updated GCR simulator capable of providing three to five consecutive mono-energetic ion beams, with rapid switching between ion species.\cite{norbury_galactic_2016} The NASA Space Radiation Laboratory (NSRL) is located at Brookhaven National Laboratory in Brookhaven, NY. Currently, NSRL is the only U.S. facility with the capapbilites to generate heavy-charged particles at energies relevent to space radiation studies. While an improvement upon previous methods, NASA's new GCR simulator remains  limited in its ability to emulate the GCR environment of deep space. The simulator lacks the capacity to generate the pions (subatomic particles) and neutrons that would follow spallation reactions, though these would make up 15-20\% of a true intravehicular dose.\cite{Slaba2015,chancellor_emulation_2017} Sequential beam exposures remain ineffective in modeling complex and simultaneous exposures of the actual GCR environment, and there is significant debate regarding the appropriate order of ion exposures delivered (as alteration of exposure sequence can affect the outcomes of an experiment).\cite{chancellor_emulation_2017,Elmore_2011} Finally, dose-rate delivered by this simulator will remain significantly higher than the radiation dose-rate anticipated for human crews during spaceflight.\cite{norbury_galactic_2016,Slaba2015}

As an additional challenge, SPE radiation has a unique dose distribution with respect to whole body irradiation. Research has demonstrated that the biological response to space radiation is unique due to a non-homogeneous, multi-energetic dose distribution.\cite{kennedy_biological_2014,romero-weaver_effect_2013} The majority of the protons in SPEs have energies less than 100MeV, with Bragg peaks that occur inside the body and LET of 10-80keV/$\mu$m (Figure \ref{fig:bragg_LET}(b)). At these energies, an exposed human would be expected to receive a much higher absorbed dose to skin and subcutaneous tissues than to internal organs.\cite{hu_modeling_2009,cengel_using_2010,coutrakon_simulation_2007,kim_evaluation_2006} Until recently, these SPE-specific toxicity profiles and dose distributions were poorly understood. As a result, the majority of prior research has been based largely on simplified models of radiation transport, relying upon simple spherical geometry to estimate organ dose approximation at average depths.\cite{billings_body_1973,wilson_issues_1995} However, with this new evidence of heterogeneous dose distribution, spherical geometry is insufficient for the modeling of radiation delivered within the space environment.

\subsubsection*{Animal Model Sensitivity and Dose Simulation}

For ease of dose specification and modeling, mono-energetic protons and GCR ions in the 100-1,000MeV range are often used for \emph{in vivo} animal model experiments such that the entire target is contained within the plateau portion of the depth-dose distribution.\cite{kennedy_effects_2008,paganetti_nuclear_2002,slater_clinical_2006,tilly_influence_2005,wambi_protective_2009} In experimental animals that are much smaller than humans, simple scaling of particle energies to match dose distribution dramatically alters the LET spectrum for the protons (Figures \ref{fig:bragg_LET} and \ref{fig:LET}). Conversely, delivering a simulated SPE or GCR exposure to smaller animals without scaling the energies would match their respective LET spectrum but create an heterogeneous dose distribution that is higher to internal organs than to superficial tissues, the exact inverse of the human SPE dose distribution.\cite{cengel_using_2010} For smaller animals (such as rodents), it is not possible to match both the LET spectrum and dose distribution of an SPE using protons.\cite{cengel_using_2010,sanzari_acute_2013,sanzari_relative_2014} Larger animal models, such as pigs or primates, allow for matching of the anticipated dose distribution for human SPE exposure using protons with a similar LET spectrum; thus, larger animal models are more likely than smaller species to provide robust estimations of human-specific space radiation effects.\cite{cengel_using_2010} However, it remains unclear whether the concurrent exposure to low-dose and dose-rate GCR radiation can be successfully emulated in small or large animal models.\cite{little_systematic_2008} Modeling of GCR radiation effects may be similarly altered by variations in animal species; however, without dedicated efforts towards expanding understanding of these phenomena, prediction of the biological consequences of long-term GCR exposure will remain theoretical at best.

Animal models pose further challenges in the development of meaningful and accurate analog research. While animal models are used in radiobiology studies as surrogates to obtain data that typically cannot be gained in ethical studies of humans, there are numerous metabolic, anatomic, and cellular differences between humans and other animal species.\cite{gawrylewski_trouble_2007} Most of the animals used in all U.S.~scientific research are mice and rats, bred specifically for use in research endeavors. While larger species are likely to provide more meaningful correlation to human effects,\cite{williams_animal_2010} due to animal protection issues and relative societal value, less than one quarter of 1\% of scientific studies are performed on non-human primates and less than one half of 1\% of studies use dogs and cats. Few studies utilize rabbits, guinea pigs, sheep, pigs, or other large mammals. While rodent experiments have contributed significantly to our understanding of mechanisms of disease, including disease caused by radiation, their value in predicting the effectiveness of treatment modalities for human application has remained controversial.\cite{hackam_translation_2006,perel_comparison_2007,hackam_translating_2007}

\begin{table}
\caption{LD$_{50}$ of various animal models used in space radiobiology studies compared to the human LD$_{50}$ dose following radiation exposures. This broad spectrum in LD$_{50}$ values emphasizes the difficulty in interpreting results of studies using specific radiation exposures in different animal models and translating them into clinical outcomes in humans. \emph{Note}: Table is adapted from the reported results of Harding 1988,\cite{harding_prodromal_1988} Morris \& Jones 1988,\cite{morris_comparison_1988} and Hall \& Glaccia 2012.\cite{hall_radiobiology_2012}}
\small 
\centering
\vspace*{0.5em}
\begin{tabular}{lcl}
\toprule
\textbf{Species}	& \textbf{LD$_{50}$},\textbf{(Gy)}	& \textbf{Reference}\\
Ferret		& $\textless$ 2 		& Harding\\
Pigs		& 2.57 			& Morris \& Jones\\
Dogs		& 2.62			& Morris \& Jones\\
Primates	&4.61			& Morris \& Jones\\
Mice		&8.16			& Morris \& Jones\\
Humans		&3--4			& Hall \& Garcia\\
\toprule
\end{tabular}
\label{tab:LD50}
\end{table}

Differences between animals and humans are clearly demonstrated by the characteristics of \emph{radiation-induced death} (RID). The \emph{LD$_{50}$} defines the required dose of an agent (\emph{e.g.} radiation) necessary to cause fatality in 50\% of those exposed. As illustrated in Table \ref{tab:LD50}, remarkably different LD$_{50}$ values have been reported for radiation exposure among different species. Currently, the genetic and physiologic basis for inter- and intra-species variation in LD$_{50}$ is not well understood. Mice have been the most extensively developed model for human diseases including radiation-induced tissue damage. Rodent models have a high potential utility in describing the physiologic and genetic basis for many aspects of the mammalian radiation response. Even so, it should be noted that, in addition to simple physiological differences between mice and larger animals (including significantly higher metabolic rate, shorter lifespan, and lower body mass), the LD$_{50}$ for mice is significantly higher than that of most other mammalian species, including humans.

It has been proposed that the differences between the LD$_{50}$ values for humans compared to small mammals, like rodents, are due to different mechanisms involved in RID at these dose levels. For mammals, death at the LD$_{50}$ dose is thought to be caused by the \emph{hematopoietic syndrome}, which includes destruction of precursor cell lines within blood-forming organs. Historically, it was thought that infection and hemorrhage are the major causes of death from hematopoietic syndrome, with one or the other of these factors predominating in different species' responses to lethal radiation exposure.\cite{lorenz1954radioactivity} For example, bacterial infection is the predominate factor leading to RID in mice at doses near their respective LD$_{50}$ levels.\cite{lorenz1954radioactivity,miller_role_1950,boone_relation_1956} However, recent results from Krigsfeld \emph{et al.} have indicated that \emph{radiation-induced coagulopathy} (RIC) and clinical sequelae that mimic \emph{disseminated intravascular coagulation} (DIC) can result in hemorrhage, microvascular thrombosis, organ damage, and death from multiorgan failure from exposure of large animals (including ferrets and pigs) to doses of radiation at or near the species' LD$_{50}$.\cite{krigsfeld_evidence_2014,krigsfeld_evidence_2014-2,krigsfeld_is_2013,krigsfeld_effects_2012,krigsfeld_mechanism_2013} RIC-associated hemorrhage occurs well before the expected decline in peripheral platelet counts after irradiation. Rodents do not exhibit signs of hemorrhage or disorders of primary hemostasis at time of necropsy after lethal radiation exposure at doses near the LD$_{50}$ dose, while large animals, including humans, do exhibit hemorrhage at death following radiation exposure. These findings suggest that humans may be at risk for coagulopathy-induced complications after radiation exposure in addition to the classically anticipated (delayed) concerns of infectious sequelae or cell-count decline, effects that may not be modeled by rodent surrogates.


\begin{figure*}[!ht]
  \begin{center}
    \subfigure[]{\label{fig:lymphocyte}\includegraphics[scale=0.7]{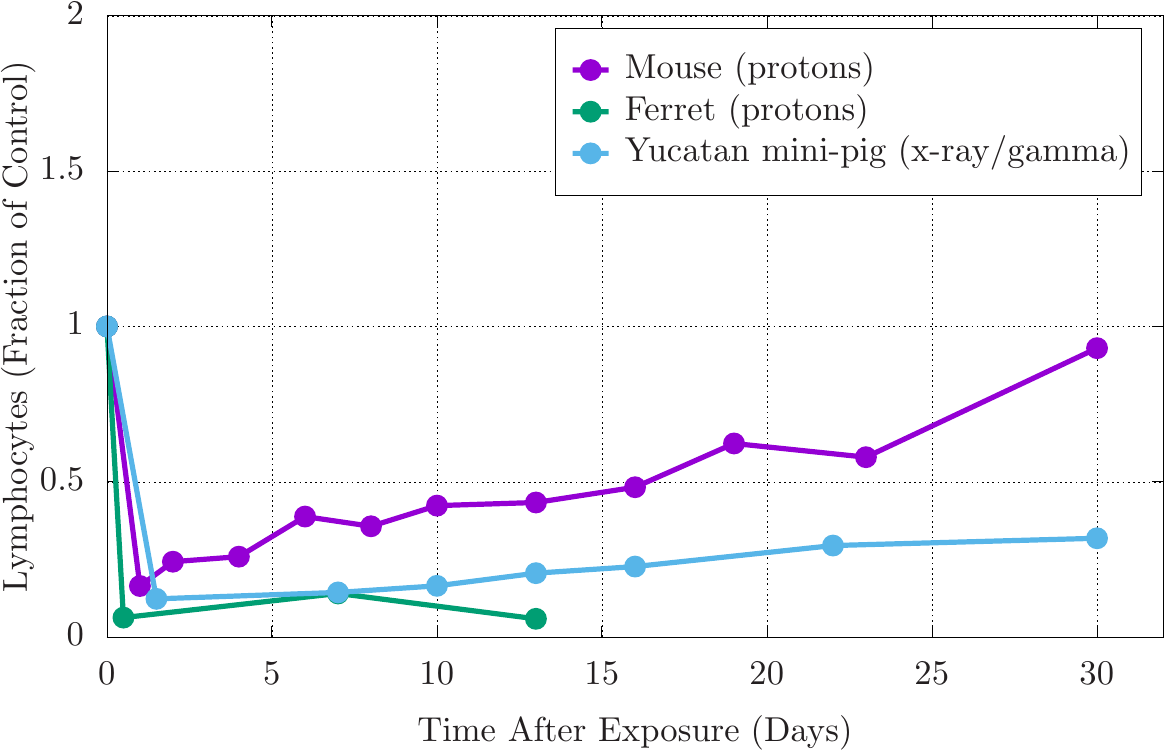}}
    \subfigure[]{\label{fig:neutrophil}\includegraphics[scale=0.7]{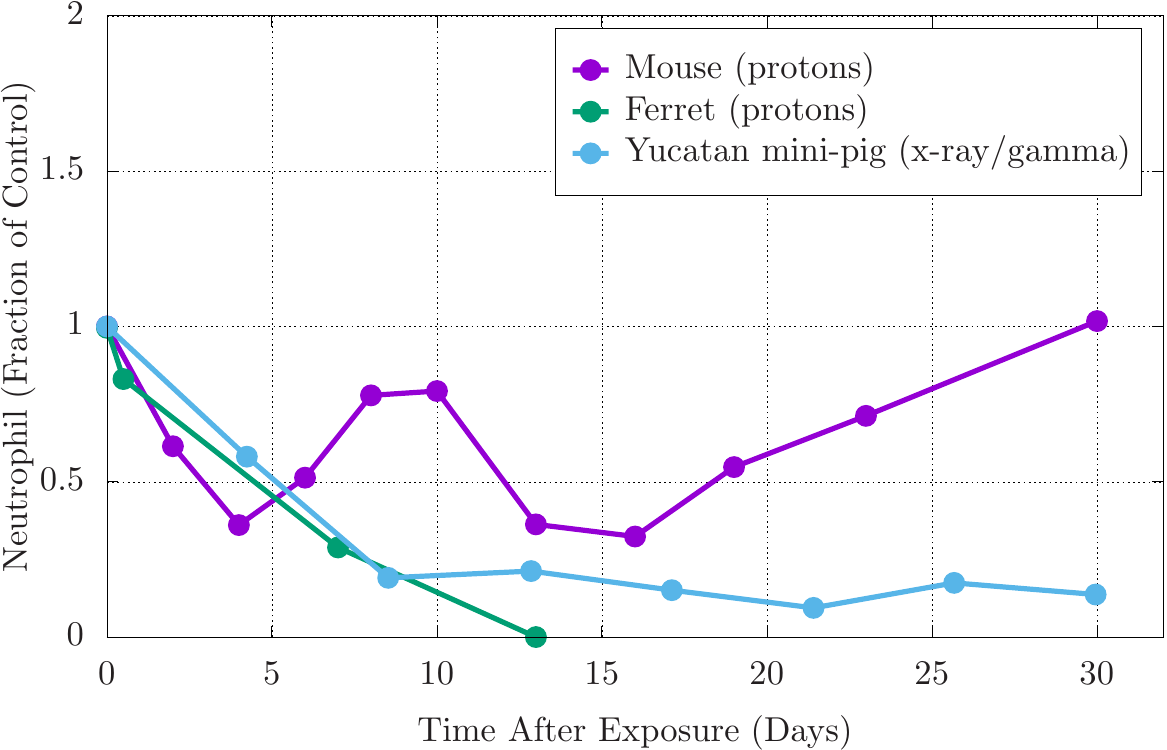}}
  \end{center}
\caption{Comparison of lymphocyte and neutrophil counts following proton and x-ray (comparable to gamma radiation) exposures in mice, ferrets, and Yucatan mini-pigs. The relative fraction of lymphocyte (a) and neutrophil (b) counts following a homogeneous proton or x-ray  exposure to the bone marrow compartment are shown. Note: calculations indicate that animals received approximately a 2Gy marrow dose. In both cases, the mouse models demonstrated the ability to fully recover within 30 days following proton exposures while the ferret and pig models showed no recovery. The ferrets were euthanized at day 13.\cite{kennedy_biological_2014} The RBE values for white blood cell counts varied greatly between the mice, ferret and pig models. RBE values were greater in ferrets than mice, and considerably greater in pigs compared to either ferrets or mice. This suggests that model-specific sensitivity to radiation exposure may lead to drastically different results in experimental outcome, leading to difficulty in extracting clinical significance from animal models with dissimilar radiation sensitivity compared to humans. Figure adapted from Kennedy\cite{kennedy_biological_2014} (mouse and ferret results) and Krigsfeld \emph{et al.}\cite{krigsfeld_evidence_2014,krigsfeld_evidence_2014-2} (Yucatan mini-pig results)}
\label{fig:mammal_WBC}
\end{figure*}

Further, RBE values for proton irradiation vary between animal models. In general, RBE values increase with animal size, with mini-pigs demonstrating higher RBEs than ferrets, and ferrets, in turn, exhibiting higher RBEs than mice (Table \ref{tab:RBE}).\cite{kennedy_biological_2014} Numerous studies have focused on RBE values for hematopoietic cells in mice at various time points after the animals have been exposed to different doses of proton or gamma radiation. \cite{maks_analysis_2011,romero-weaver_effect_2013} In these rodent models, RBEs do not differ significantly from 1 at any of the time points or doses of radiation evaluated. However, similar studies in ferrets and mini-pigs have demonstrated alterations of RBE value that are dependent upon animal model, type of radiation, time since exposure, and cell-line evaluated (for example, total white blood cell count vs. neutrophils). In one study, proton-irradiated ferrets examined 48h after exposure demonstrated RBEs for white blood cells ranging from 1.2-1.6 and RBEs for neutrophils ranging from 1.9-2.1.\cite{sanzari_effects_2013} In Yucatan mini-pigs evaluated 4 days after exposure, the RBEs for white blood cells was found to be 2.4-4.1 and the RBEs for neutrophils was 2.2-5.0 (see Table \ref{tab:RBE}, Figure \ref{fig:mammal_WBC}).\cite{sanzari_relative_2014}

In other experiments, proton exposure in mini-pigs again resulted in significantly greater hematopoietic injury and white blood cell count reduction than comparable gamma exposure (Figure \ref{fig:pig_LD50}).\cite{sanzari_acute_2013,sanzari_relative_2014} The results of these studies demonstrate that RBE values of different radiation types, calculated for the same endpoints, can vary greatly by animal species and cell line. One contributing factor may be the repair capacity of the blood cell renewal systems in mice; such capabilities appear to be lacking in mini-pigs (an animal model with more human-like hematopoietic characteristics), making them more susceptible to radiation-induced declines in cell counts. Given the presumed closer approximation of radiation effects in larger animals to human-specific consequences, this suggests that space radiation-specific RBE values for humans may be considerably higher than those in mice.
\begin{table}
\caption{The Relative Biological Effectiveness (RBE) for SPE-like protons compared with standard reference radiations (gamma or electron) in animal models.\cite{sanzari_effects_2013,sanzari_relative_2014,maks_analysis_2011,romero-weaver_effect_2013} The RBE of proton exposure varies greatly for total white blood cells (WBC) and specifically for neutrophils when comparing animal models. Note that ferret RBE values were determined 48h after exposure; mini-pig values were determined 4d post-irradiation.}
\small 
\centering
\vspace*{0.5em}
\begin{tabular}{lclll}
\toprule
\textbf{Animal} &	& \textbf{WBC} &	& \textbf{Neutrophil} \\
Mouse		&		& 1 		 &		& 1\\
Ferret		&		& 1.16--1.6 &		& 1.9--2 \\
Mini-Pig 	&		& 2.4--4.1	 & 		& 2.2--5\\
\toprule
\end{tabular}
\label{tab:RBE}
\end{table}

These studies demonstrated novel efforts towards an integrated, physiology-based approach for the evaluation of organ system- and species-specific endpoints. Using a more comprehensive evaluation of radiation toxicity for multiple doses and dose-rates in multiple animal models, this effort advanced the understanding of the impact of genetic heterogeneity and demonstrated that animal model, physiology, body mass, and fidelity of a space radiation analog (in this case, a multi-energy proton spectrum) all contribute to radiation response. Such efforts towards the integration of the numerous factors that contribute to radiation-induced effects will be critical to translation of research results and prediction of clinical responses in humans.

\begin{figure}[ht]
\centering
\includegraphics[width=0.95\columnwidth]{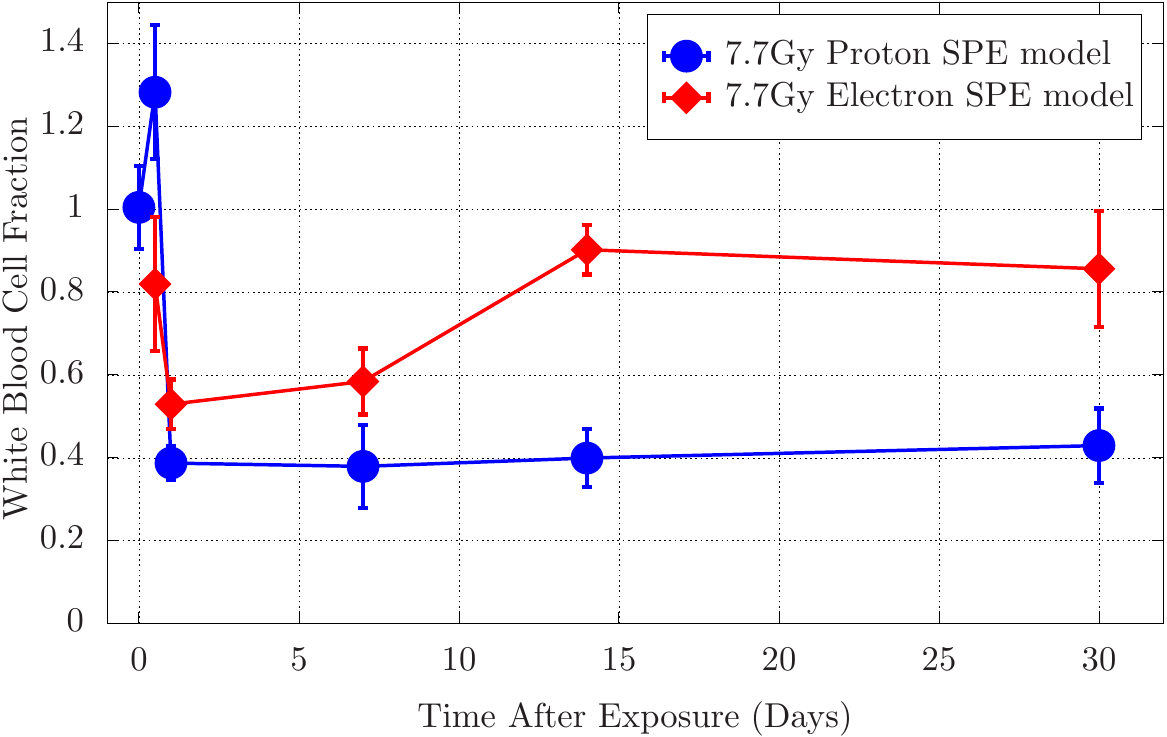}
\vspace*{1.5em}
\caption{Results from Yucatan mini-pigs exposed to simulated Solar Particle Event (SPE)-like radiation consisting of several different energies of protons. In this study, Kennedy \emph{et al.} utilized an inhomogeneous distribution of protons that resembled a SPE spectrum, as demonstrated in Figure \ref{fig:bragg_LET}. Electrons were used as the surrogate radiation for determining the RBE following exposure to a SPE-like distribution of protons. Electrons were chosen because a SPE-like distribution could not be achieved with $^{60}$Co as demonstrated in Figure \ref{fig:89SPE}. Note the white blood cell counts in the mini-pig model recovered to near pre-irradiation levels following exposure to the electron radiation while the white blood cell counts for those exposed to a SPE-like proton spectrum remained suppressed for 30 days after exposures. These results indicate that the mini-pigs were not capable of repairing the hematopoietic damage caused by the proton radiation exposure as efficiently as they could repair the electron radiation damage. Figure adapted from Kennedy 2014.\cite{kennedy_biological_2014}}
\end{figure}

Finally, studies of the synergistic effects of radiation combined with spaceflight environment stressors (\emph{e.g.} microgravity, environmental factors, isolation and emotional stress, etc.) show that such factors in combination impart an increased susceptibility to infection and delayed wound healing.\cite{maks_analysis_2011,sanzari_combined_2011,wilson_comparison_2012} While spaceflight medical capabilities have been developed for the management of some acute injuries, such as wound care and infection control, it is unclear whether standard management techniques will be effective against the synergistic variables that alter wound healing and associated risks specific to the space environment. Historically, there has been limited testing on the efficacy of management techniques, including pharmaceutical interventions, when radiation exposure is a factor. Similarly, few research protocols examining operational medical care have included the additional variables of the high-stress and isolated environment,\cite{levine_immunosuppression_1998,crucian_terrestrial_2014,pagel_effects_2016} infections related to the altered bacterial and chemical exposures specific to space vehicles,\cite{alekhova_diversity_2015,pierson_microbial_2001,norbiato_microbial_1998,mermel_infection_2013} or factors related to gravitational unloading,\cite{sanzari_combined_2011,li_broad-spectrum_2015,li_hindlimb_2014,zhou_effect_2012} and no studies have effectively examined all of these variables simultaneously. It is unclear whether these complex interactions can be fully simulated even in large animal models for appropriate extrapolation of human risk. There is a need to better understand the mechanism of the synergistic effects observed, define appropriate animal models for analog research efforts, and determine efficacy of standard treatments against damage resulting from radiation-combined injury. Dedicated effort towards these goals will better allow for operationally relevant and appropriate countermeasures.\cite{dicarlo_medical_2008}

\subsubsection*{Translation of Space Radiobiology Research to Human Health Outcomes}

Biological damage from radiation exposure is generally classified as \emph{deterministic}, dose threshold-based effects related to significant cell damage or death (for example, the spectrum of clinical manifestations that make up Acute Radiation Sickness), or \emph{stochastic}, where increased exposure is associated with increased risk though no threshold dose is necessary for biological impact (for example, carcinogenesis).\cite{national_council_on_radiation_protection_and_measurements_potential_2011} Currently, carcinogenesis is the only long-term, stochastic effect that has a clearly defined permissible exposure limit in spaceflight. Terrestrial radiation (\emph{e.g.} occupational or clinical radiotherapy gamma or x-ray exposures) is known to be associated with carcinogenic risk;\cite{beir_health_2006} at this time, there is no definitive evidence that space radiation causes human cancer, but it is reasonable to assume that it can. The dose-equivalent of radiation received by astronauts currently traveling to the ISS for 6 months is approximately 100mSv;\cite{shavers_implementation_2004} doses of 100mSv of terrestrial radiation sources have been associated with an elevated cancer risk in human populations.\cite{beir_health_2006} NASA's "Lifetime Surveillance of Astronaut Health" (LSAH) program documents cancer cases in astronauts, among other health parameters. Previous review of LSAH data suggests that there may be evidence of increased cancer risk in astronauts compared to a control population, though data are inconclusive and limited by the very small sample size.\cite{institute_of_medicine_u.s._review_2004}

Most evidence for the effects of space-like radiation exposures in humans has been derived from epidemiological studies on the atomic-bomb survivors, radiotherapy patients, and occupationally-exposed workers. These studies have focused on the association between ionizing radiation exposure and the long-term development of degenerative tissue effects such as heart disease, cataracts, immunological changes, cancer, and premature aging for moderate to high doses of low-LET radiation.\cite{ncrp_guidance_1989,national_council_on_radiation_protection_information_2006} The findings are further supported by results of laboratory studies using rodent animal models.\cite{blakely_review_2007} However, true risks for these diseases from low dose-rate exposures to GCR and intermittent SPE are much more difficult to assess due to long latency periods and the numerous challenges involved in studying the radiation environment.\cite{blakely_review_2007} Additionally, the types of radiation exposure produced by atomic bombs (high dose and high dose-rate gamma and neutron radiation) are dissimilar to radiation exposures for astronaut crews during spaceflight.

The theoretical, calculated RBEs for some space radiation-induced cancers are quite high, which has led to speculation that the risk of cancer development from space radiation exposure is at least as high, and perhaps higher, than the risk of cancer development from exposure to radiation on Earth.\cite{kennedy_countermeasures_2011,cucinotta2017} However, there are currently no biophysical models that can accurately project all acute, subacute, degenerative, and carcinogenic risks specific to the range of particles and energies of ionizing radiation in the space environment. There is little information available about dose response and dose-rate modifiers for specific effects or about the degenerative effects associated with ionizing radiation, and very few biological models describe degenerative processes (e.g. cardiovascular degeneration) caused by ionizing radiation.\cite{huff_risk_2009}

Exposure to the LEO radiation environment has been associated with alterations to chromatin structure.\cite{Bender1967,Fedorenko2003,Testard1997,george2011} However, it is not well understood how such damage relates to impacts on cellular function or long-term carcinogenic risk. There is a paucity of understanding regarding the interpretation of chromosomal damage rates identified in astronauts and the long-term effects induced by the space radiation environment, without relying on terrestrial studies of different radiation sources, doses, dose-rates, or complexity for context. For example, NASA's Human Research Program Evidence Report on the Risk of Radiation Carcinogenesis\cite{HRPRiskCarcinogen2016}, published in 2016, cites numerous studies to provide an assessment of risk for chromosomal damage (and, ultimately, carcinogenesis). A review of the studies cited in this report highlights the limitations described throughout this manuscript, including reliance upon mono-energetic radiation sources\cite{Durante2002,Belli2002,george2011,Hada2007,Loucas2015,Wang2014}, comparison to or interpretation of results in the context of gamma or x-ray exposures\cite{Belli2002,Hada2007,Johannes2004,Loucas2015,Wang2014}, or use of dose or dose-rates far exceeding those expected during spaceflight.\cite{Durante2002,george2011,Loucas2015,Wang2014} Indeed, many of these same factors are cited as limitations to NASA's primary radiation cancer risk prediction model.\cite{CucinottaRisk2012}

In addition, few studies have assessed mutation rates due to LEO radiation at a whole genome level. Whole genome sampling techniques are being employed for other carcinogenic stressors.\cite{alexandrov_mutational_2016} Direct observations of mutation rates, as well as an understanding of the epigenetic changes and cellular damage using \emph{in vitro} cell culture models, may now be possible due to recent advances in long-term cell culture aboard the ISS.\cite{sharma_personal_2016} Quantification of observable mutation rates from LEO exposures may better inform future modeling efforts and provide a critical understanding of the molecular mechanisms behind observed pathologies. However, even data obtained from the LEO environment is less than ideal, as the ISS is heavily shielded and the close proximity of the Earth provides significant protection from radiation exposure. While improved understanding of the LEO environment may help inform risk predictions, there is significant work to be done in characterizing these risks in the radiation environment \textit{outside} of LEO.

\section*{Discussion}

The health risks associated with exposures to space radiation will become more onerous as future manned spaceflight missions require extended transit outside of LEO and beyond the protection of the Earth's magnetosphere. The indigenous shielding provided by the Earth's magnetic field attenuates the major effects of space radiation exposures for current LEO missions; in the highly mixed-field environment of interplanetary space, radiation dose could increase dramatically. Even behind shielding, secondary particles produced by interactions of primary cosmic rays and the atomic molecules of the spacecraft structure can deliver a significant fraction of the total dose equivalent. Astronaut crews could be exposed to multiple SPEs of unpredictable magnitude with doses that could induce clinical illness and exacerbate biological outcomes from the chronic GCR environment.

The limited accumulation of knowledge to date has yet to provide sufficient data for even an estimation of total risk, let alone predictions of human clinical outcomes or appropriate mitigation strategies before, during, or after exposure. Accurately simulating the spectrum of energies, ion species, doses, and dose-rates found in the space radiation environment is a non-trivial endeavor. For the numerous reasons described above, emulation of the radiation environment, choice of surrogate animal model, and delivery of appropriate complexity, rate, and magnitude of dose have all limited the knowledge available for extrapolation of radiation risk within the context of spaceflight. These factors have limited our ability to develop operational and useful medical countermeasures to mitigate the radiation risk of future exploration-class spaceflight.

To improve upon the limitations described, there must be a focused effort to develop novel or new methods of simulating the space radiation environment in more realistic analogs. This should include more realistic dose-rate studies that can determine if presumed or modeled outcomes are being observed at mission relevant dose-rates and dose. Additionally, heavier utilization of the animal laboratory on board the ISS with comparison of tissues, organ, and blood samples, identifying realistic dose thresholds and dose-rates, and comparing these data to ground-based studies, would greatly improve the current approach to analog construction. The use of animal models should be strategic and consistent with species, strain, dose, and dose-rates with an effort towards the highest-fidelity studies possible for human risk extrapolation.\cite{williams_animal_2010} While rodent models may be highly useful for initial characterization studies and for statistically significant outcomes, true advances are more likely to come from an effort to utilize larger animals with more human-like physiology for landmark studies on how specific outcomes may translate to humans. Finally, while there would be numerous challenges and ethical considerations involved, studies of non-human primates for final validation of risk and mitigation strategies would likely prove highly beneficial for the protection of future human crews.

\begin{figure}[!ht]
\centering
\includegraphics[width=0.95\columnwidth]{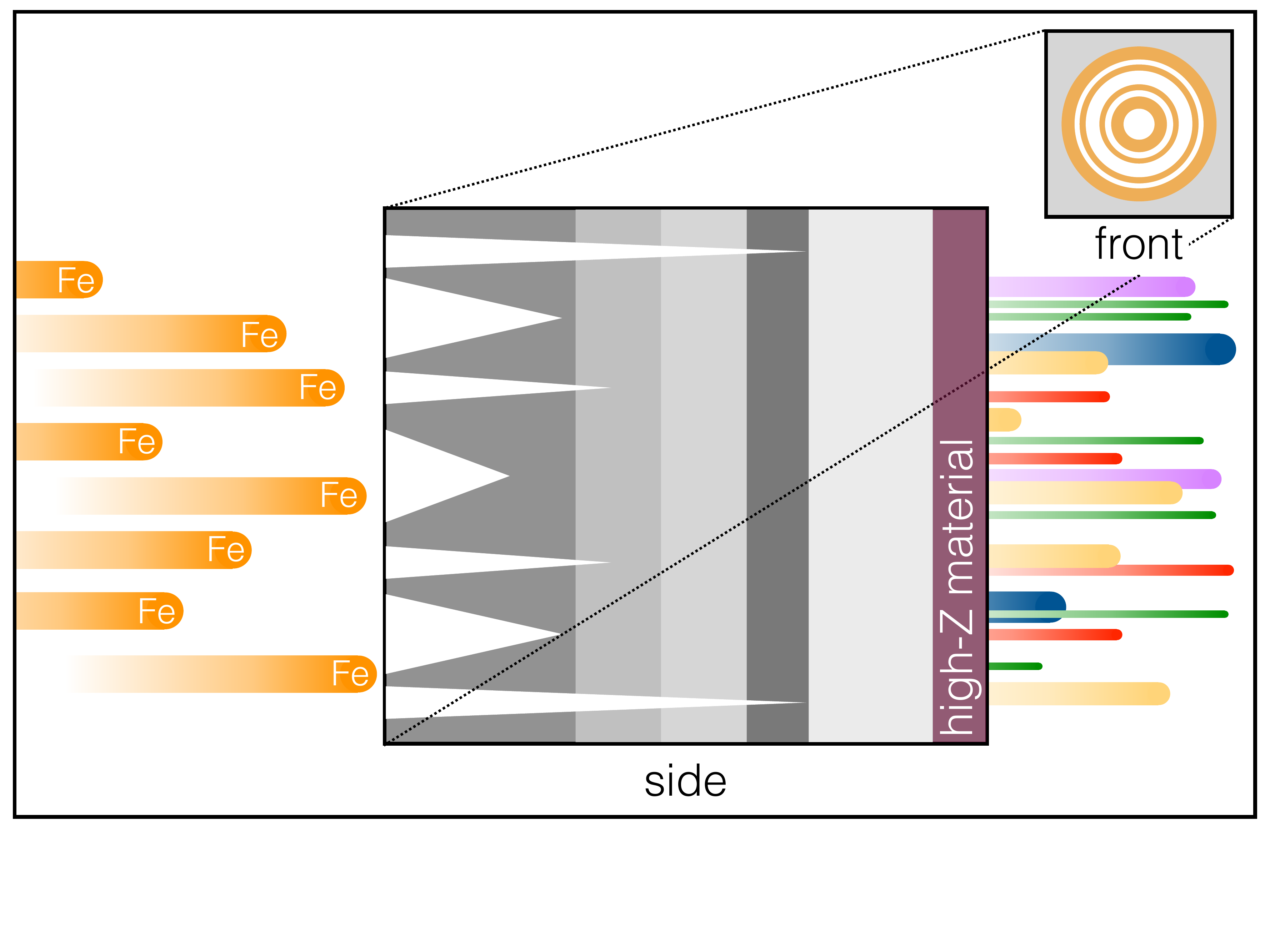}
\vspace*{2.5em}
\caption{Moderator block geometry concept for the emulation of space radiation spectra. A primary beam of $^{56}$Fe (iron, left) is selectively degraded with a carefully-designed moderator block to produce a desired distribution of energies and ions (represented by the colorful lines on the right) simulating the intravehicular space radiation environment. To preferentially enhance fragmentation and energy loss, cuts (white sections on the left) are performed in the moderator block made up of different materials (depicted by different shades of gray). Before the spallation products exit the moderator block, a high-Z material layer is added for scattering. The inset shows the circular beam spot, as well as the symmetric cuts made into the moderator block. Adapted from Chancellor \textit{et al.}~2017.\cite{chancellor_emulation_2017}}
\label{fig:blockart}
\end{figure}

As described above, NASA's updated GCR simulator may be able to provide some improvements to simulation studies by use of rapid-sequential mono-electric beam exposures.\cite{norbury_galactic_2016,Slaba2015} Recent developments by Chancellor \emph{et al.}~demonstrate the potential for more accurate analog recreation of the GCR radiation environment by allowing for continuous generation of ionizing radiation that more closely matches the ion distribution, LET spectrum, and dose-rate of GCR (Figure \ref{fig:blockart}).\cite{chancellor_emulation_2017} These recent findings suggest that the radiation environment inside spaceflight vehicles can be experimentally generated by perturbing the intrinsic properties of hydrogen-rich crystalline materials in order to produce specific nuclear spallation processes when placed in an accelerated mono-energetic heavy ion beam. Such an approach could allow for vast improvements to the simulation of the complex mix of nuclei and energies found in the space radiation spectrum.\cite{chancellor_emulation_2017}

Potential radiation exposure to astronaut crews occurs on a timescale that is measured in days to months for SPE and GCR. Technological, practical, and financial considerations make continuously irradiating animals for more than a few hours exceedingly difficult. In addition, because the lifespan of most experimental animals is more than an order of magnitude shorter than the human lifespan, the interpretation of long-term, low dose-rate exposures using such models would be questionable even given the open opportunity to perform long-duration experiments. As radiation dose-rate can have a major impact on modulating the severity of the radiation response, it is critical to obtain at least some dose-rate data for radiation experiments investigating clinical outcomes of space radiation exposures. While some radiation effects are either unchanged or mitigated by decreased dose-rates, data on non-targeted radiation effects (such as genomic instability and adaptive responses) suggest that dose response could be altered at lower dose-rates, with significant differences in quantitative (slope of the dose-toxicity curve) or qualitative (toxicity effects) biological responses. This is especially true for high-LET radiation exposure under conditions of increased oxidative stress promoted by spaceflight.\cite{azzam_ionizing_2012,rizzo_effects_2012,buonanno_long-term_2011,matsumoto_new_2009} In previous studies on SPE-like radiation, dose-rates from 17cGy/hour up to 50cGy/minute have been modeled experimentally and statistical analysis of these data have begun to explore the potential quantitative or qualitative impact of dose-rate on the toxicity of multi-energy spectrum.\cite{chancellor_space_2014} Use of such data to better design dose-rate extrapolation experiments would be highly useful for more robust, future studies.

There have been other advances in fields related to space radiation effects, including whole genome sequencing as well as transcriptional, proteomic, and epigenomic studies of cellular response. There is a growing list of genes known to affect radiation sensitivity for many different biological effects of radiation (\emph{e.g.} molecular, chromosomal, signal transduction-associated growth-regulating changes, cell killing, animal tissue and tumor acute and late effects, and animal carcinogenesis). Even so, there is a need to correlate observed sequence changes with corresponding alterations of radiosensitivity.\cite{national_council_on_radiation_protection_and_measurements_potential_2011} Incorporation of these investigational directions opens new opportunities to evaluate space radiation risk on a genomic level, defining risk and allowing for improved understanding of the pathology of radiation-induced injury and the potential for intervention in such processes.

Finally, there are a number of lessons that may be learned from historical spaceflight and the health of early space pioneers, though it has been difficult to extract meaningful conclusions from historical data. For example, some sources suggest that there is no statistically significant increase in carcinogenesis in Apollo, Space Shuttle, or ISS astronaut crews in comparison to the average U.S. population; other reviews of data suggest that risk is indeed increased for astronauts.\cite{chancellor_space_2014,institute_of_medicine_u.s._review_2004,cucinotta2017,ncrp_guidance_1989} Given that the broad research base has utilized non-ideal and highly-limited analogs for the prediction of risk, the fact that reality has deviated from theoretical, calculated risk is not entirely surprising. Medicine does not advance without clarifying treatment options using human subjects. Models and animal data are useful surrogates for space radiation studies but provide limited benefit for the interpretation to human outcomes, and studies on humans exposed to occupational radiation and clinical radiotherapy are imperfect proxies. The reliance upon these surrogates continues to limit the ability to translate radiation knowledge to spaceflight scenarios.

We now have the benefit of a larger, cumulative astronaut population that has flown in space while exposed to a variety of doses that exceed the identified thresholds for some degenerative and carcinogenic outcomes. The health of these astronauts, including early indicators of disease, is closely monitored by NASA medical and epidemiological resources with yearly medical examinations and careful records of clinical outcomes. This provides critical, real human data that could be used to evaluate the actual long-term health risk of space radiation. Understandably, these data are limited to highly sensitive and protected internal review in order to ensure the privacy of flown astronauts, given small sample sizes and the risk of inadvertent identification through mission- or demographic-specific details. This should not preclude NASA from taking advantage of these data points while remaining vigilant with prioritizing the privacy and protection of astronaut medical health records. The application of this source of data will enhance our understanding of the true risk of space radiation, the characterization of human clinical outcomes, and the development of appropriate mitigation strategies.

\section*{Conclusions}

The scientific community has struggled to collect meaningful and robust data for the characterization of the space radiation environment and the risk that such an environment poses to future astronaut crews. While many of the challenges outlined herein have plagued historical research endeavors, there are significant improvements that could be made to research design that would improve upon our ability to better predict risk and provide realistic strategies and risk posturing for future exploration spaceflight. Use of improved modeling techniques to emulate the space environment, selection of appropriate biological surrogates for extrapolation of human effects, and careful use of flown astronaut data could provide much-needed advances in space radiation research. As humans seek to explore space outside of the close proximity and protection of LEO, we have the responsibility to address the space radiation risk to the extent of terrestrial capabilities in order to provide the best information and protection possible for our future explorers.

\section*{List of Acronyms}{The following abbreviations are used in this manuscript:\\}
\newline
\noindent
cGy: centiGray\\
cm: centimeter\\
CNS: Central Nervous System\\
Co: Cobalt\\
DIC: Disseminated Intravascular Coagulation\\
g: gram\\
GCR: Galactic Cosmic Ray\\
GeV: Giga Electron Volt\\
Gy: Gray\\
HZE: High-Charge \& High-Energy\\
ISS: International Space Station\\
J: Joule\\
keV: Kilo Electron Volt\\
kg: kilogram\\
LD$_{50}$: Lethal Dose for 50\% of exposed population\\
LET: Linear Energy Transfer\\
LEO: Low Earth Orbit \\
LSAH: Lifetime Surveillance of Astronaut Health\\
MeV: Mega Electron Volt\\
mGy: milliGray\\
mSv: milliSievert\\
NASA: National Aeronautics \& Space Administration\\
NCRP: National Council on Radiation Protection \& Measurements\\
RBE: Relative Biological Effectiveness\\
REID: Risk of Exposure-Induced Death\\
RIC: Radiation-Induced Coagulopathy\\
RID: Radiation-Induced Death\\
SPE: Solar Particle Event\\
Sv: Sievert\\
$\mu$m: micrometer\\
$W_{\rm R}$: Weighting Factor\\

\vspace*{0.5em} \noindent {\color{bv} \bf Acknowledgments}

\noindent {\scriptsize \begin{spacing}{1.00}
H.G.K.~acknowledges support from the NSF (Grant No.~DMR-1151387). Part of the work of H.G.K.and J.C.C.~has been based upon work supported by the Office of the Director of National Intelligence (ODNI), Intelligence Advanced Research Projects Activity (IARPA), via Interagency Umbrella Agreement IA1-1198. The views and conclusions contained herein are those of the authors and should not be interpreted as necessarily representing the official policies or endorsements, either expressed or implied, of the ODNI, IARPA, or the U.S. Government. The U.S. Government is authorized to reproduce and distribute reprints for Governmental purposes notwithstanding any copyright annotation thereon.
\end{spacing}}

\vspace*{0.5em} \noindent {\color{bv} \bf Author Contributions} \noindent {\scriptsize \begin{spacing}{1.00}

J.C.C. developed the concept of the review. J.C.C., K.A.C., and H.G.K. contributed to the discussion on space physics. J.C.C., R.S.B., S.M.A. and K.A.C. contributed to the discussion on operational space radiation. J.C.C., K.A.C. and A.R.K. contributed to the discussion on dosimetry. J.C.C., R.S.B., S.M.A., K.H.R., and A.R.K. contributed to the discussion on countermeasures. R.S.B., S.M.A., K.A.C., K.H.R., and A.R.K. contributed to the discussion on clinical effects of space radiation on humans. J.C.C., R.S.B., S.M.A., K.A.C., K.H.R, and A.R.K. contributed to the discussion on space radiobiology. J.C.C., R.S.B., S.M.A., K.A.C., K.H.R, and A.R.K. contributed to the discussion on animal models. R.S.B., K.H.R., and A.R.K. contributed to the discussion on genetics. J.C.C., K.A.C., and H.G.K. contributed to the discussion on computational modeling. All authors contributed equally to the review of the literature, discussion on the interpretation of research outcomes to spaceflight operations, and drafting of the manuscript.

\end{spacing}}

\vspace*{0.5em} \noindent {\color{bv} \bf Competing financial interests}

\noindent {\scriptsize The authors declare that they have no competing
financial interests.}

\end{document}